\documentclass[12pt,preprint]{aastex}
\usepackage{graphicx,lscape}
\newcommand\ea{\it et al. \rm}

\newcommand\ros{{\sl ROSAT }} 
\newcommand\cha{{\sl Chandra}} 

\newcommand\xmm{XMM-{\sl{Newton}}} 
\newcommand\ergsec{\hbox{erg s$^{-1}$}}
\newcommand\ergcm{\hbox{erg cm$^{-2}$ s$^{-1}$}}

\shorttitle{Extended emission around PSR B0355+54}
\shortauthors{Tepedelenlio\v{g}lu \& \"{O}gelman}

\begin{document}
  \title{\bf Discovery of extended emission around the pulsar B0355+54}
  
  \author{E. Tepedelenlio\v{g}lu\altaffilmark{1}
    \email{emre@cow.physics.wisc.edu} 
    \and
    H. \"{O}gelman\altaffilmark{1,2}}
  \email{ogelman@cow.physics.wisc.edu} 
  \altaffiltext{1}{Department of Physics, University of
    Wisconsin-Madison, 1150 University Ave., Madison, WI 53703, USA}
  \altaffiltext{2}{Faculty of Engineering and Natural Sciences,
    Sabanc\i University, Orhanl\i Tuzla, \.{I}stanbul 34956, Turkey}

  \begin{abstract}
  PSR B0355+54 is one of the handful of pulsars that has been observed
  with both \cha\ and \xmm. The analysis of the archival data has
  revealed the pulsar and a $\sim30$\arcsec\ compact nebula
  surrounding it. \xmm\ also has detected a trail which extends
  $\sim$6\arcmin\ and similar to the compact nebula is also
  counter-aligned with the proper motion of the pulsar. The spectrum
  of both the pulsar and the extended emission are well described by
  an absorbed power-law model. The measured flux corresponds to an
  efficiency of converting the spin-down luminosity into X-rays in the
  2-10 keV band of $\sim$0.01\% and $\sim$1\% for the pulsar and
  the extended emission, respectively. From the \xmm\ data we have
  detected pulsations at the expected radio period. The energetic and
  the extent of the extended emission can be explained by a Bow-shock
  formed by the motion of the pulsar through the interstellar medium.
    
  \end{abstract}
  
  \keywords{pulsars:individual (\objectname{PSR B0355+54}) ---
    stars:neutron --- X-rays:stars}

  \section{Introduction}
  
  PSR B0355+54 is a middle age ($\tau\sim$0.56 Myr) pulsar with a
  spin-down energy loss rate of $\dot{E}=4.5\times 10^{34}$ \ergsec,
  and a relatively close distance of $d=1.04^{+0.21}_{-0.16}$ kpc
  \citep{chatterjee}. It was observed with \ros\ for a total of 19 ks
  and was detected with a non-thermal luminosity (2-10 keV) of
  $1.2\times 10^{31}$ \ergsec\ \citep{slane}. Taking into account the
  extended emission discussed in this paper, one should consider this
  as only an upper-limit to the pulsar's X-ray emission, considering
  \ros\ does not have a sufficient angular resolution to resolve these
  features.

  Pulsar wind nebulae (PWNe) are useful probes of pulsars and their
  surroundings. In the case of older ($\tau>10^{5}$ years) isolated
  neutron stars such as Geminga the PWN is thought to be the Bow-shock
  created by the pulsar proper motion through the interstellar medium
  (ISM) \citep{deluca,caraveoa}. The distance, the energy loss rate,
  and the proper motion of some of the pulsars is well known through
  radio observations. The knowledge of such parameters make pulsar
  Bow-shocks particularly appealing to study. In this letter we report
  the discovery of an extended emission associated with the pulsar
  B0355+54, and investigate the possible implications.

  \section{Observations}\label{observations}
  \subsection{\cha}\label{chandra}
  
  PSR B0355+54 was observed using the Advanced CCD Imaging
  Spectrometer (ACIS) instrument aboard the \cha\ X-ray Observatory on
  2004 July 16. Data were collected in the nominal timing mode, with
  3.241 s exposures between CCD readouts, and "FAINT" spectral
  mode. The standard \cha\ screening criteria produced a total usable
  exposure time of 67.13 ks.  The ACIS image reveals a point source
  near the pulsar radio position and a relatively faint diffuse
  emission that extends $\sim30$\arcsec\ towards the south-east side,
  which we label as the compact nebula (CN).  The detected point
  source is located at $\alpha$=$03^{\mbox{h}}58^{\mbox{m}}53\fs72$,
  $\delta$= +54\arcdeg13\arcmin13\farcs9 (J2000.0) which is only
  0\farcs2 away from the proper motion corrected pulsar radio
  position. This source is detected with an ACIS-S countrate of
  0.0031$\pm$0.0002 cps in the 0.2-10 keV band. The errors quoted in
  this work are in the 1$\sigma$ confidence range, unless otherwise
  noted.
    
  After removing the contributions from the point sources, we smoothed
  the image to examine the extended emission. The resulting image is
  given in Figure~\ref{smoothimage}. As can be seen, there is excess
  emission towards south-east of the pulsar.  We compared the surface
  brightness of this extended emission to that of the background's and
  calculated the countrate of the extended emission from an annulus
  situated at the pulsar position that extends from 3\arcsec to
  35\arcsec. The background countrate was measured from an annulus
  that extends from 37\arcsec\ to 70\arcsec. This annulus was chosen
  to be co-centric with the extraction region used for the extended
  emission. The observed extended emission gives an excess of
  714$\pm$69 counts over the background. This corresponds to a net
  surface brightness of 670$\pm$61 counts arcmin$^{-2}$. We also
  compared the background used for this analysis to a region without
  any sources taken away from the pulsar, towards the chip
  boundary. This resulted in no discrepancies between the surface
  brightness for the two regions.

  \subsection{\xmm}\label{xmm}
      
  PSR B0355+54 was observed with \xmm\ on 2002 February 10 for 29 ks
  and 28 ks with EPIC-MOS and EPIC-pn instruments, respectively. The
  EPIC-pn EPIC-MOS1, and EPIC-MOS2 instruments were operated in the
  small-window, full-window (imaging), and timing modes which provide
  a temporal resolution of 6 ms, 2.6 s, and 1.5 ms, respectively. The
  medium filters were used for MOS, the thin filter was used on
  pn. The EPIC-pn and -MOS1 cameras detect a point a source consistent
  with the \cha\ and the radio position.

\begin{figure}
    \includegraphics[width=16cm]{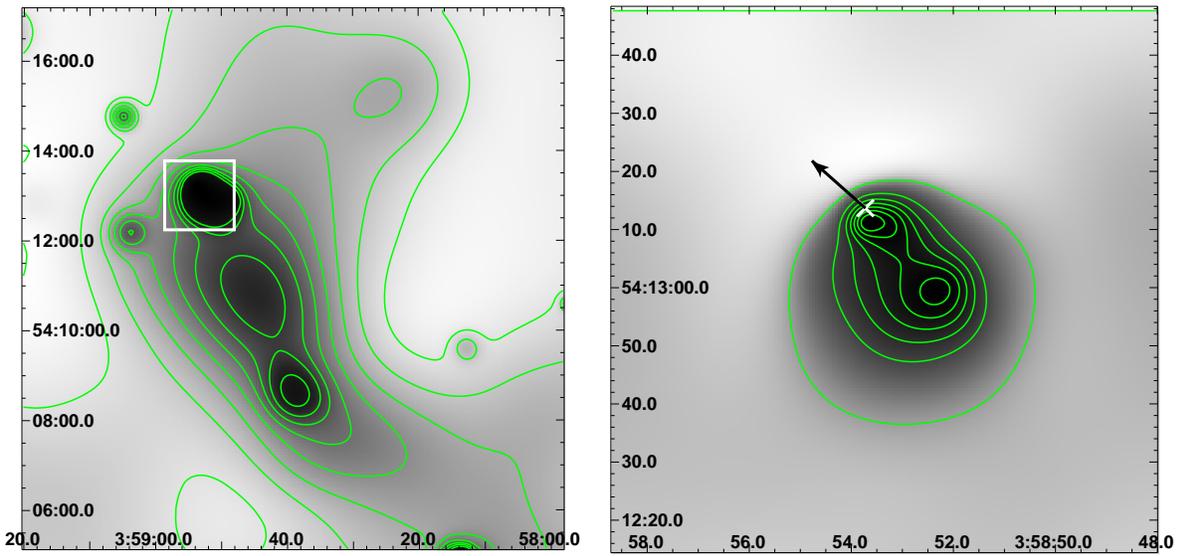}
    \caption{Gray-scale plots of the 0.2-10 keV image centered on
      the pulsar PSR B0355+54 ({\it{left}}) by \xmm\ EPIC-MOS and
      ({\it{right}}) \cha\ ACIS. The X-ray point source contribution
      is removed, the images are smoothed with a Gaussian of width
      2\arcsec. Also shown are the isophoto contours. On the left
      panel the box shows the zoomed in region shown on the right
      panel (\cha\ image). The direction of the pulsar proper motion
      is shown on the right panel with a black arrow and the pulsar
      position is indicated by a white cross ($\times$). The length of
      the proper motion vector is the distance the pulsar would travel
      in 1000 years.}
    \label{smoothimage}
  \end{figure}

  The CN detected with \cha\ is also visible in the \xmm\ data (see
  Figure~\ref{smoothimage}). Although \xmm\ data has resolved the
  extended emission it is difficult to perform any spectral analysis
  on the CN due to relatively low spatial resolution. Consequently,
  this region in EPIC-MOS data (similar for EPIC-pn) contains
  significant amount of counts from the pulsar. For example, in an
  annular region that is centered on the source of interest and
  extends from 10\arcsec-35\arcsec\ contains $\sim$35\% of the pulsar
  counts.

  We should also note that in the \xmm\ data only there is extended
  emission that extends up to $\sim 7$\arcmin\ that is in the opposite
  direction of the proper motion, we label this feature as the trail
  (see Figure~\ref{smoothimage}). This structure is aligned with the
  CN and at an initial glance seems fainter. To quantify this, we
  extracted counts from an elliptical region that is aligned with the
  pulsar's proper motion and is 3\farcm1 wide and 5\farcm8 long which
  was located 1\farcm5 south-east of the pulsar. This region contained
  316$\pm$48 background-subtracted source counts corresponding to a
  surface brightness of 22.7 counts arcmin$^{-2}$. When we analyzed
  the Chandra data no such emission was detected. This is not
  surprising since \cha\ is less sensitive with respect to \xmm. For
  example, the \cha\ observation of Geminga pulsar was unable to
  confirm the two tails detected with the \xmm\ observatory
  \citep{deluca,pavlov}.
  
  \section{Timing}

  PSR B0355+54 has a period of 0.156 s. When \cha\ ACIS is operated in
  imaging mode its temporal resolution is 3.2 seconds which is not
  sufficient to resolve the periodicity of the pulsar. When operated
  in the small window mode \xmm\ EPIC-pn camera has a timing
  resolution of 6 ms. This becomes 1.5 ms for the EPIC-MOS2 which was
  operated in the timing mode, where imaging is made only in one
  dimension to obtain a higher temporal resolution.

  In order to check for any pulsations we first extracted source
  counts from a circle centered on the pulsar with a radius
  20\arcsec. Such a small radius was used to reduce the background due
  to the unpulsed extended emission surrounding the pulsar. The
  extracted event times were barycenter corrected. We then calculated
  the value of the $Z_{1}^{2}$ \citep{buccheri} at the predicted
  pulsar frequency (see Table~\ref{tab1}). The obtained value of 7.9
  corresponds to a probability of chance occurrence of 0.019. We also
  applied the same criteria to the EPIC-MOS2 data. Since the majority
  of the data in the vicinity of the pulsar is below 0.2 keV we took
  all photons with energies below 2 keV.  The $Z_{1}^{2}$ for this set
  at the expected frequency was 17.5, which has a probability of
  0.00016 of occurring by chance.

\begin{figure}
    \includegraphics[width=11cm,angle=-90]{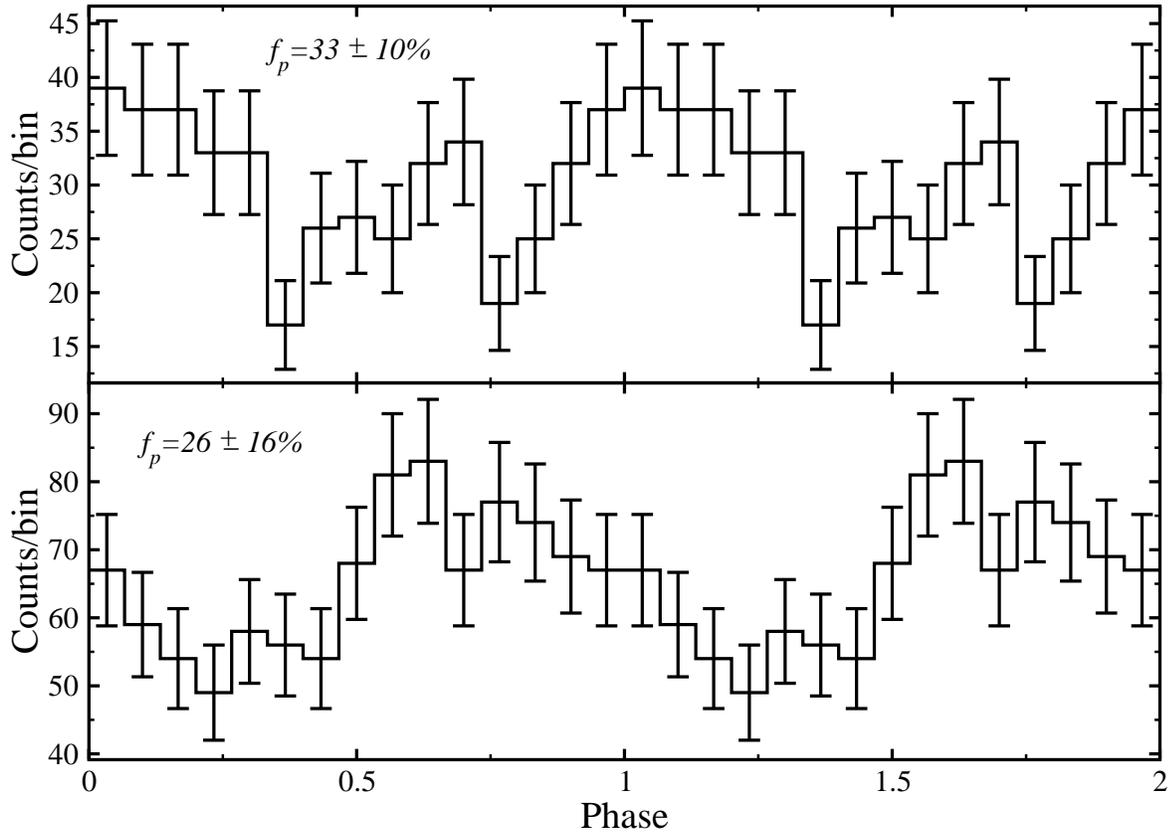}
    \caption{X-ray pulse profiles of PSR B0355+54, constructed from
      EPIC-MOS2 {\it{top}} and EPIC-pn {\it{bottom}}. The pulse
      fractions are given in the figure. Notice the double-peak nature
      of the pulse profiles. The peaks are $\sim$0.3 of a period
      apart. Two cycles are shown or clarity.}
    \label{pf}
  \end{figure}

  At the radio period we constructed the pulse profiles for both MOS2 and
  pn event times (Figure~\ref{pf}). Both profiles show a double peak
  nature, however the smaller peak is not significant, specifically in
  the MOS2 data. As was discussed by \citet{beckera} this kind of a
  pulse profile would not be observed from a neutron star emitting
  from its hot polar caps. Together with spectral data this is
  suggestive of magnetosphere being the main source of X-ray emission
  for PSR B0355+54.

  \section{Spectral  analysis}
  \subsection{Pulsar}
  For the \cha\ observation the pulsar's energy spectrum was extracted
  from a 2\arcsec\ circular region centered on the pulsar
  position. Whereas the background was extracted from an annular
  region extending from 37\arcsec\ to 70\arcsec. The reason for
  selecting the background away from the pulsar is due to the extended
  emission surrounding the immediate vicinity of the pulsar. The
  source region contains $\ga$95\% of the source counts.

  The spectrum extracted from the MOS1 data was done by selecting all
  events detected in a circle of radius 20\arcsec\ centered on the
  pulsar position. Using the \xmm\ EPIC-MOS model point spread
  function (PSF), 75\% of all source counts are within this
  region. The reason for selecting such a small source region, and in
  turn fractional encircled energy, was to minimize the contamination
  due to the events originating from the extended emission. The
  background spectrum for the MOS1 data was extracted from an annular
  region centered on the pulsar position which extends from
  55\arcsec-200\arcsec. For the EPIC-pn data we also used an
  extraction radius of 20\arcsec\ centered on the pulsar.  This
  selection radius includes 75\% of the point-source flux.  Since the
  point source was close to the edge of the CCD, we extracted the
  background spectrum from a source free region around the pulsar from
  a circular region with a radius of 45\arcsec. The extracted spectra
  for PSR B0355+54 was binned so as to have 15, 25 and 20 counts per
  bin, for ACIS, EPIC-pn and EPIC-MOS, respectively.

  Among the single-component spectral models fitted to \cha\ spectra a
  power-law model was found to give the best ($\chi^{2}$=7.6 for 9
  degrees of freedom [dof]) representation of the observed energy
  spectrum of the pulsar. A single blackbody model for the pulsar gave
  neither a statistically nor a physically acceptable fit
  ($\chi^{2}$=21.9 for 9 dof). The parameters for the best fit
  power-law model are given in Table~\ref{tab2}. The value obtained
  for the absorbing column is consistent with the value derived
  through the basic assumption that there is 10 neutral hydrogen atoms
  for each free electron, giving a value of $N_{\mbox{H}}\sim 2\times
  10^{21}$ cm$^{-2}$. The measured X-ray luminosity corresponds to an
  efficiency of converting the spin-down luminosity into X-rays in the
  2-10 keV band of 0.009\%.

  The EPIC-MOS and -pn spectra were fit simultaneously to get the best
  fit parameters. As was done for \cha\ spectrum we tried the
  single-component models, power-law and blackbody. It was found that
  absorbed power-law model gave the best representation of the data
  (see Table~\ref{tab2}). Single blackbody model did not give a
  statistically acceptable ($\chi^{2}$=51.6 for 33 dof) spectral
  fit. We also fitted an absorbed power-law model all pulsar spectra
  (ACIS, EPIC-MOS and -pn) simultaneously. This fit yielded a higher
  photon index but the best values for the parameters are consistent
  with single \cha\ and simultaneous \xmm\ fits (see
  Table~\ref{tab2}).

  \subsection{Extended emission}
  The spectrum for the CN was selected from an annular region
  co-centric with the source region that has inner and outer radii
  3\arcsec\ and 35\arcsec, respectively. The background for this
  spectrum was chosen to be the same as the pulsar's spectrum. For the
  analysis of the trail we extracted the spectrum from the region
  described in $\S$~\ref{xmm}. The background for this spectrum was
  extracted from an off-source region south-west of the trail.

  The extracted spectra for the CN and the trail were binned so as to
  have 25 and 50 counts per bin, respectively. We limited the
  generated spectra to the energy interval 0.2-10 keV. The absorbed
  power-law model gave a statistically acceptable fit with
  $\chi^{2}$=24.7 for 42 dof for the CN and $\chi^{2}$=15.9 for 22 dof
  for the trail. Table~\ref{tab2} lists the best fit parameter values
  for the extended emission. The measured values given in
  Table~\ref{tab2} for the luminosities of both the CN and the tail
  account for only 0.74\% of the pulsar spin-down. We also tried a
  thermal bremsstrahlung model to fit the trail. This model also gave
  a statistically acceptable fit with a $\chi^{2}$=16.1 for 25
  dof. The characteristic temperature of the region is
  $kT=6.5^{+18.3}_{-3.0}$ keV. The absorbing column is consistent with
  that of the power-law fit, $N_{\mbox{H}}=4.4^{+2.2}_{-1.4}\times
  10^{21}$ cm$^{-2}$. The resulting emission measure ($\int
  n_{e}n_{H}dV$) is $\sim 1.9\times 10^{54}$ cm$^{-3}$, at a distance
  of 1.04 kpc.

  \section{Discussion}
  The \cha\ and \xmm\ observations of PSR B0355+54 have proven the presence of extended emission associated with the pulsar.
  This emission has two components which are $\sim$30\arcsec\ and
  $\sim$6\arcmin\ long (see $\S$~\ref{observations}). The recently
  detected proper motion direction (see Table~\ref{tab1}) being
  counter aligned with the trail together with the observed spectrum
  being well fitted with a power-law model suggests that the observed
  extended emission may be the result of synchrotron radiation from a
  wind-driven nebula supported by the ram pressure created by the
  proper motion through the ISM. This hypothesis have been found to be
  consistent with the size and energetics of the X-ray trails of
  Geminga \citep{caraveoa} and PSR B1929+10 \citep{beckera}.

  There are two relevant time scales that need to be considered. The
  first is the $\tau_{flow}$, the timescale for the passage of the
  pulsar over the length of its X-ray trail. The other one is
  $\tau_{syn}$, the synchrotron lifetime of the radiation. This is
  given by the equation \citep[see e.g.][]{gaensler}
  $$ \tau_{syn} =39 B_{\mu G}^{-3/2} \left( \frac{h\nu}{\mbox{keV}}
    \right)^{-1/2} \mbox{kyr},
  $$ where $B_{\mu G}$ is the magnetic field in the
  emission region in units of $\mu$G. The first one of these can be
  estimated through the relation $\tau_{flow}=r_{t}/v_{p}$, where
  $r_{t}$ is the extent of the trail which is measured to be
  $\sim6$\arcmin. This at a distance of 1.04 kpc corresponds to a
  linear distance of 1.8 pc.  The proper motion, $v_{p}$, is measured
  by \citet{chatterjee} and is 61$^{+12}_{-9}$ km s$^{-1}$. This,
  together with the linear extent of the trail, gives the time it took
  PSR B0355+54 to travel across the trail as $\sim 34$ kyr. Assuming
  these two time scales are comparable the estimate for the magnetic
  field around the emitting region is $B_{\mu G}\sim$1 for an X-ray
  photon of energy $h\nu=1$ keV. We should note that this
  calculation ignores the possible inclination of the pulsar's proper
  motion with respect to the line of sight.

  Assuming that the ISM in the vicinity of the pulsar carries a
  magnetic field that does not differ significantly from that of the
  average of our galaxy which is $\sim 4$ $\mu$G \citep{beck}. This
  value should be the characteristic magnetic field for the trail
  where the X-ray emitting particles are no longer in the shocked
  region. For the CN, where the electrons are in the shocked region,
  we need to consider the change in density due to the shock and hence
  the change in the magnetic field. We can use standard hydrodynamical
  arguments to show that the density in the shock region is given by
  the relation
  $\rho_{\mbox{\small{shock}}}=4\rho_{\mbox{\small{ISM}}}$, which
  would imply a magnetic field of
  $B_{\mbox{\small{shock}}}=4B_{\mbox{\small{ISM}}}$ or
  $B_{\mbox{\small{shock}}}=16$ $\mu$G, given the average ISM magnetic
  field. This value, given the large uncertainties, is in agreement
  with the value found here.

  The termination shock radius is given by the balance of the ram
  pressure between the wind particles and the ISM,
  i.e. $$\frac{\dot{E}}{4\pi c R_{s}^{2}}=\frac{1}{2}\rho v_{p}^{2},$$
  where $\rho$ is the density of the ISM and $v_{p}$ is the proper
  motion velocity. From here one can estimate the radius of the
  termination shock as $R_{s}\sim3\times
  10^{6}\dot{E}_{34}^{1/2}n^{-1/2}v_{p,100}^{-1}$ cm, where $n$ and
  $v_{p,100}$ are the number density in the ISM in units of cm$^{-3}$
  and the pulsar space velocity in units of 100 km s$^{-1}$,
  respectively. Adopting $n=1$ cm$^{-3}$ the shock termination radius
  becomes 1.1$\times 10^{17}$ cm. This corresponds to an angular size
  of $\sim$7\arcsec\ at a distance of 1.04 kpc. This value is
  consistent with the data given the observational constraints.

\begin{figure}
    \includegraphics[width=12cm,angle=-90]{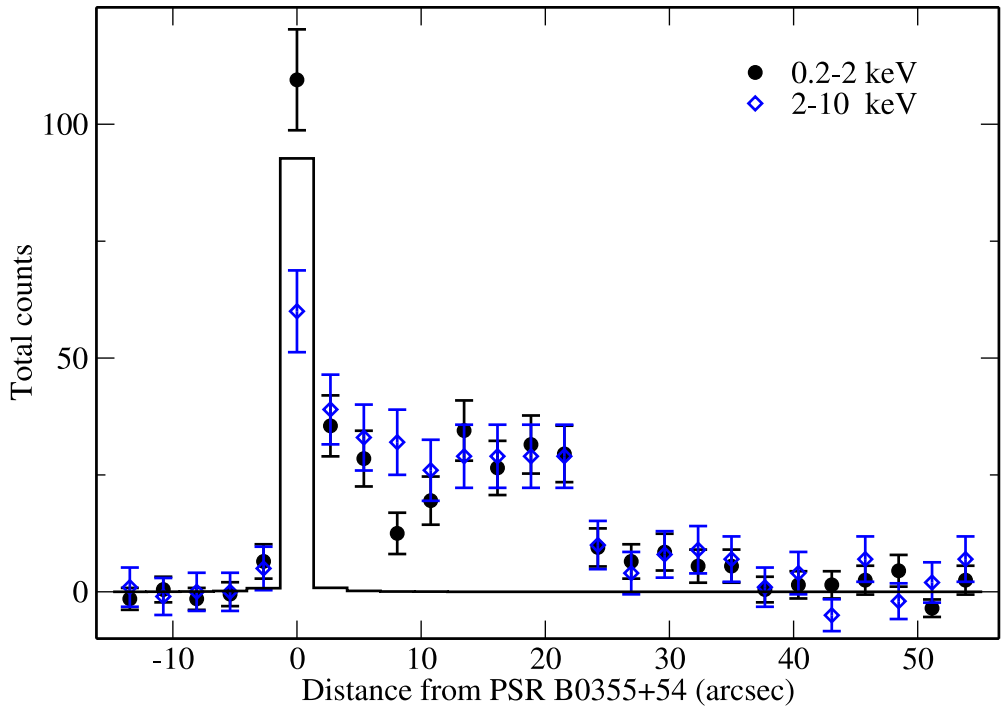}
    \caption{The background-subtracted linear profile of the CN, along
    the proper motion direction as seen by \cha\ in the energy range
    0.2-2 keV ({\it{filled circles}}) and 2-10 keV
    ({\it{diamonds}}). The solid line is the contribution from the
    pulsar B0355+54 in the 0.2-2 keV range.}
    \label{radial}
  \end{figure}

  In order to check the energy dependence of the extent of the CN we
  constructed two background-subtracted linear profiles along the
  proper motion direction, which contain photons from the intervals
  0.2-2 keV and 2-10 keV (Figure~\ref{radial}). For comparison also
  the point source contribution is plotted. Although the structure of
  the CN is not identical in both energy ranges the overall extent
  within the statistical uncertainties are equal. This is consistent
  with the identification of this region as the termination
  shock. This region is expected to show a constant X-ray spectrum
  across its extent. Due to the pressure difference between the
  regions ahead and behind the pulsar's motion the termination shock
  is not of uniform radius around the pulsar and is extended along the
  proper motion direction \citep{bucciantini}. The larger structure
  seen by \xmm\ (trail) can be thought of as the emission originating
  from the wind particles that flow around the edge of the shocked
  region. The magnetic field in this region should be an order of
  magnitude less than the nebular magnetic field in the shocked
  region, if equipartition holds. The two components of the extended
  emission could be explained by the Bow-shock theory.

  PSR B0355+54 is yet another pulsar with known proper motion and a
  collimated outflow in the opposite direction. Such flow, also
  observed in Geminga, could be interpreted as a pulsar jet
  \citep{pavlov}. This is reminiscent of the argument proposed for the
  Vela pulsar \citep{markwardt} where the outflow in opposite
  direction is proposed as the mechanism to boost pulsar velocities in
  contrast to birth-kicks.

  \clearpage
  
  \begin{deluxetable}{ll}
\tablewidth{0pc}
\tablecaption{Properties of PSR B0355+54 derived from radio observations
\label{tab1}}
\tablehead{Parameter & Value}
\startdata
Frequency (Hz).........................................& 6.39458383 \\
Frequency derivative (Hz s$^{-1}$)..................&-1.797906$\times 10^{-13}$\\
Epoch (MJD)............................................&52315.6\\
Spin-down age ($10^6$ yr)............................& 0.56\\
Spin-down energy ($10^{34}$ ergs s$^{-1}$).............& 4.5\\
Inferred Magnetic Field ($10^{11}$ G).............& 8.4\\
Dispersion Measure (pc cm$^{-3}$).................& 57.2\\
Distance (kpc)..........................................& 1.04\\
\multicolumn{2}{c}{Position (J2000)} \\
$\alpha$ ..............................................................& $03^{\mbox{h}}58^{\mbox{m}}53\fs72$\\
$\delta$ ............................................................... & +54\arcdeg13\arcmin13\farcs9 \\
\multicolumn{2}{c}{Proper motion (mas yr$^{-1}$)} \\
$\mu_{\alpha}$ .............................................................. & 9.20$\pm$0.18\\
$\mu_{\delta}$ .............................................................. & 8.17$\pm$0.39\\

\enddata
\tablecomments{Distance, position, and proper motion are all taken from 
\citet{chatterjee}.}
\end{deluxetable}

  \clearpage
  
  \clearpage
  
  \clearpage
  
  \begin{deluxetable}{lccccc}
\tablewidth{0pc} 
\tablecaption{Spectral fit results for PSR B0355+54.
\label{tab2}}

\tablehead{
\colhead{Comment}&\colhead{$\chi^{2}$/(dof)} &
\colhead{$\Gamma$} & \colhead{$N_{\mbox{H}}$} &
\colhead{$f_{\mbox{x}}^{2-10}$} & \colhead{$L_{\mbox{x}}^{2-10}$} \\
\colhead{}&\colhead{}&\colhead{}& \colhead{10$^{21}$cm$^{-2}$} &
\colhead{10$^{-14}$\ergcm}& \colhead{10$^{31}$\ergsec}}

\startdata 
\multicolumn{6}{c}{\cha} \\
\tableline
Pulsar & 7.6/9&1.24$^{+0.16}_{-0.24}$ & 0.70$^{+1.68}_{-0.70}$&3.11$^{+0.40}_{-0.43}$& 0.40\\
CN & 24.7/42 & 1.44$^{+0.26}_{-0.34}$ & 6.10$^{+1.75}_{-1.84}$ & 12.3$^{+1.9}_{-4.3}$\tablenotemark{a} & 1.61\\
\tableline
\multicolumn{6}{c}{\xmm\tablenotemark{b}} \\
\tableline
pn+MOS & 37.3/32 & 1.54$^{+0.26}_{-0.15}$ & 4.99$^{+2.62}_{-1.50}$&11.3$^{+2.0}_{-3.9}$ & 1.47\\ 
pn+MOS+ACIS& 50.7/43 & 1.41$^{+0.10}_{-0.12}$ & 3.32$^{+1.00}_{-1.17}$ &3.14$^{+0.43}_{-0.58}$& 0.41\\
Trail & 15.9/22 & 1.84$^{+0.49}_{-0.36}$ & 5.43$^{+3.37}_{-1.72}$ & 13.5$^{+2.5}_{-5.0}$ & 1.72
\enddata

\tablecomments{The quoted errors are 1$\sigma$}
\tablenotetext{a}{The value quoted is the integrated flux in the
annular region where the spectra for the extended emission was
extracted.}

\tablenotetext{b}{The first and second row tabulate the best fit parameters
for the spectral fit performed simultaneously on data extracted from
EPIC-pn and -MOS and  EPIC-pn, -MOS, and ACIS, respectively. The third row shows the values for the absorbed power-law best fit parameters.}

\end{deluxetable}

  \clearpage

\end{document}